# Nanoscale Plasmonic and Optical Modulators Based on Transparent Conducting Oxides

**Zhaolin Lu\*, Wangshi Zhao, and Kaifeng Shi**

*Microsystems Engineering, Kate Gleason College of Engineering,*

*Rochester Institute of Technology, Rochester, New York, 14623, USA*

[*]Corresponding author: zhaolin.lu@rit.edu

## Abstract

Recent experiments showed that unity-order index change in a transparent conducting oxide (TCO) can be achieved in a metal-oxide-semiconductor (MOS) structure by accumulation charge. However, the ultrathin (~5nm) accumulation layer and inherent absorption of TCOs impede the practical applications of this effect. Herein, we propose and explore a novel waveguide, namely "TCO-slot waveguide", which combines both the tunable property of a TCO and field enhancement of a slot waveguide. In particular, light absorption can be sharply enhanced when the slot dielectric constant is tuned close to zero. Based on TCO-slot waveguides, efficient electro-absorption modulation can be achieved within 200 nm with small insertion loss.

Ultra-compact high-speed electro-optic (EO) modulators have become one of the critical technical bottlenecks impeding the wide applications of on-chip optical interconnects. This is due to the very poor electro-optic (EO) properties of conventional materials[1]. Phase modulators are on the order of millimeters[2-5] and can reduce to about tens of micrometers by introducing novel structures to enhance the EO effect[6,7]. Absorption modulators can be quite compact, but in most cases they require advanced materials [8-12]. Even though, their dimensions are still 10$\mu$m~100$\mu$m. On-chip optical interconnects require EO modulation at the nanoscale.

The key to achieve nanoscale EO modulation is to (1) identify an efficient and low cost active material and (2) greatly enhance light-active medium interaction based on a novel waveguide or platform. Recently, we found that light absorption can be greatly enhanced in epsilon-near-zero-slot structure even when the slot width is less than 1nm [13]. In that case, graphene works as a tunable epsilon-near-zero (ENZ) material. We found an ENZ material has many advantages as an EO material: (1) sharply enhanced absorption can be achieved in an ultrathin slot; (2) the ultrathin slot does not introduce a large insertion loss; (3) an ENZ material often has tunable optical properties because a small change in carrier density may result in a significant change in dielectric constant.

The ENZ effect can be found in almost any material at $\omega \approx \omega_p/\sqrt{\varepsilon_\infty}$ according to the Drude model for dielectric constant, $\varepsilon = \varepsilon_\infty - \frac{\omega_p^2}{\omega(\omega+j\gamma)}$, where $\varepsilon_\infty$ is the high frequency dielectric constant, $\omega_p$ is the plasma frequency, and $\gamma$ is the electron damping factor. For example, |$\varepsilon$(tungsten)|=0.483 at $\lambda_0$=48.4 nm, and |$\varepsilon$(aluminum)|=0.035 at $\lambda_0$=83 nm[14]. However, the plasma frequencies of most metals are located in the ultraviolet regime due to their extreme high carrier concentration. Note $\omega_p = \sqrt{\frac{Ne^2}{\varepsilon_0 m^*}}$, depending on carrier concentration $N$, and the effective electron mass $m^*$. To shift the plasma frequency into the near infrared (NIR) regime for telecom applications, the carrier concentration should reduce to $10^{20} \sim 10^{21}/cm^3$, which coincides that of transparent conducting oxides (TCOs). Their well-known representatives, indium tin oxide (ITO) and indium zinc oxide (IZO), are degenerately doped semiconductors widely used as transparent electrodes in displays. The exploration of TCOs as the plasmonic metamaterial for NIR applications can date back to decades ago [15-19]. Comparative studies can be found in Refs. [20-22]. Feigenbaum and coworkers [23] have experimentally showed that unity-order index change in a TCO can be achieved in a MOS structure by voltage-induced accumulation charge. They have summarized their measurements as Table 1 in Ref. 23. We paid special attention to the two cases in the table: at $V_a=0$, $\varepsilon_\infty=4.55$, $\omega_p=2.0968\times10^{15}$ rad/s, $\gamma=724\times10^{12}$ rad/s [24], and $N_1=1.0\times10^{21} cm^{-3}$; at $V_a=1V$, $\varepsilon_\infty=4.37$, $\omega_p=3.4687\times10^{15}$ rad/s, $\gamma=52.9\times10^{12}$ rad/s, and $N_2=1.65\times10^{22} cm^{-3}$. Note the accumulation carrier density is determined by multiple parameters and the applied voltage is only one of them. Thus, we refer to these two cases by their corresponding carrier concentrations $N=N_1=1.0\times10^{21} cm^{-3}$ and $N=N_2=1.65\times10^{22} cm^{-3}$, which are directly related to the dielectric constant of ITO. Based on the Drude model and the measured parameters, Fig. 1(a) plots the dielectric constant of ITO accumulation layer (real part and imaginary part) as a function of wavelength under $N=N_1$ and $N=N_2$, respectively. In particular, when $N=N_1$, at $\lambda_0=1136nm$  $\varepsilon_1 = 3.2074+j0.5867$  ➔epsilon-far-from-zero state;

when $N=N_2$, at $\lambda_0=1136nm$  $\varepsilon_2 = -0.0014+j0.1395$  ➔epsilon-near-zero state.

Note that the magnitude of the dielectric constant has changed $|\varepsilon_1|/|\varepsilon_2|=23.4$ times by only 1.0V gate voltage across the 100-nm oxide! In this sense, ITO and other TCOs may be excellent EO materials. Indeed, TCOs have been proposed as the active media in several plasmonic modulators [25-27]. However, there seems always a tradeoff between the dimensions and insertion loss of the modulators. In this work, we show that modulators with ultracompact dimensions and small insertion loss can be achieved even on a plasmonic platform based on our recent work on ENZ-slot waveguides [28].

A low carrier concentration (or a voltage between 0 and 1.0V in Ref. [23]) should result in ENZ accumulation layer at the telecom wavelengths. However, due to the lack of the experimental data, we only consider devices work at 1136nm. To circumvent the band edge absorption of Si at 1136nm in the theoretical and numerical study, the waveguide semiconductor is assumed to be a material with a similar refractive index of Si at telecom wavelengths, e.g. GaAs. The applications of the devices described below can be easily extended into telecom Si photonics.

Figure 1(b) illustrates the structure to be discussed in this paper. We will use ITO as one example of various TCOs. Assume 10 nm thick ITO film is sandwiched between two Au slabs with 30nm thick $SiO_2$ buffer layer. The structure is simply the MOS structure as reported in Ref. [23] with a thinner oxide layer. Optically, it is also known as a metal-insulator-metal (MIM) plasmonic waveguide [29], where a well confined transverse magnetic (TM) plasmonic mode can be excited between the two Au slabs. The magnetic field is parallel to the slabs, i.e. $H=H_x$ in Fig. 1(b). At the $SiO_2$-ITO interface, the continuity of normal electric flux density $\varepsilon_{ITO}(E_{ITO})_y = \varepsilon_{SiO_2}(E_{SiO_2})_y$ is applicable, where the free charge effect is included in the complex dielectric constant. Thus, very high electric field can be excited when $|\varepsilon_{ITO}| \to 0$. In other words, an ENZ-slot can sharply enhance the electric field in the slot. Without loss of generality, we assume the dielectric constant of ENZ-slot to be $\varepsilon = \varepsilon' + j\varepsilon'' = \varepsilon' + \frac{j\sigma}{\omega\varepsilon_0}$. The dissipation power density $p_d = \frac{1}{2}\sigma E^2 \propto \frac{1}{2}\varepsilon'' E^2 \propto \frac{1}{2}\varepsilon''/|\varepsilon|^2$ can be greatly enhanced when $|\varepsilon| \to 0$. The absorption of the

ENZ-slot may even be many times than that of Au in the waveguide as can be seen in the following context.

Based on the transfer matrix method, we solved the TM mode supported by the Au-ITO-SiO$_2$-Au stack, i.e. a 2D ITO-slot MIM plasmonic waveguide. The dielectric constant of Au is -63.85+j5.07 at $\lambda_0$=1136nm. We considered two cases: (1) without a gate voltage, $N = N_1$ and the 10-nm ITO layer has dielectric constant $\varepsilon_1$ = 3.2074+j0.5867; (2) with a *suitable* gate voltage, $N = N_2$ and the 10-nm ITO layer is split into two, namely 5-nm unaffected layer with $\varepsilon_1$ = 3.2074+j0.5867 and 5-nm accumulation layer with $\varepsilon_2$ =-0.0014+j0.1395. As shown in Fig. 2(b), the electric field can be greatly enhanced in the accumulation layer at $\lambda_0$=1136nm when carrier concentration increases from $N_1$ to $N_2$. In particular, the magnitude of $E_y$ increases about 9.2 times. In addition, similar level of enhancement can be achieved when the ENZ-slot is sandwiched in a dielectric waveguide. Figure 2(b) shows the mode profiles of an ENZ-slot dielectric waveguide at $N_1$ and $N_2$. The top and bottom dielectric layers, each 125nm thick, are assumed to be heavily doped semiconductor with refractive index 3.45.

These 2D film stacks can be easily rendered into 3D rib waveguide as shown in Fig. 3. We used a 3D mode solver to study their modes based on the FDTD method. Figure 3(a) shows the mode profiles of the ITO-slot plasmonic waveguide at different carrier concentrations. Note the top Au strip is only 200nm wide. As can be seen, there is a considerable shift in the effective index: 1.99 at $N=N_1$, and 1.09 at $N=N_2$. Thus, quite compact phase modulators may be realized. More importantly, there is a huge change in the waveguide attenuation. At $N=N_1$, the $|E_y|$ in the ITO is even lower than in the SiO$_2$ buffer layers, and the waveguide works at the low loss state with $\alpha_1$=2.92dB/$\mu$m; at $N=N_2$, the $|E_y|$ in the accumulation layer is many times higher than in the SiO$_2$ buffer layers, and the waveguide works at the high absorption state with $\alpha_2$=23.56dB/$\mu$m. As a result, modulation depth 20.64dB/$\mu$m can be achieved, and 3dB-modulation depth only requires 146nm propagation distance! Based on the film stack shown in Fig. 2(b), a dielectric modulator can be designed. Figure 3(b) shows the mode profiles of the ITO-slot dielectric modulator at different carrier concentrations. A similar modulation effect can be achieved. The dielectric modulator may find more practical applications.

To evaluate the insertion loss of the EO modulators, we performed 3D FDTD simulations with the smallest mesh size down to 0.5nm. We first simulated the modulator based on the plasmonic waveguide platform as shown in Fig. 4(a). We assume the modulator is embedded in a waveguide with same configuration as itself except without the ITO layer. The length of the EO modulator is 150nm. Figures 4(b, c) show the power distribution in the waveguide at $N=N_1$ and $N=N_2$, respectively. Simulation results demonstrate that the overall throughput is 89.6% at $N=N_1$, and 40.8% at $N=N_2$. Note that the insertion loss is only 0.48 dB (89.6%). The achievable modulation depth, 3.42 dB, is very close to the one predicted by the 3D mode solver.

We also simulated the modulator based on the dielectric waveguide platform as shown in Fig. 4(d). The length of the EO modulator is 200nm. In this case, we assume the modulator is embedded in a dielectric waveguide with same overall dimensions as itself except without the ITO and buffer layers. Figures 4(e, f) show the power distribution in the waveguide at $N=N_1$ and $N=N_2$, respectively. Simulation results demonstrate that the overall throughput is 88.2% at $N=N_1$, and 39.1% at $N=N_2$. Note that the insertion loss is only 0.55 dB (88.2%). The achievable modulation depth, 3.53 dB, is smaller than the one predicted by the 3D mode solver. This is due to the mode mismatch between the slot waveguide of the modulator and its input/output rib waveguide. We expect that its performance (modulation depth and insertion loss) can be significantly improved by replacing the input/output rib waveguide with a dielectric slot waveguide.

In addition, the optical bandwidth of the modulators can be over several THz due to the slow Drude dispersion. The EO modulators can potentially work at an ultra-high speed, being mainly limited by the RC delay imposed by electric circuits.

To summarize, recent experiments show the potential applications of TCOs as tunable ENZ materials. When sandwiched in a plasmonic or dielectric waveguide, a very thin ENZ film can greatly enhance light absorption. The tunable ENZ-slot waveguides may enable EO modulation at nanoscale and an optical modulator can be made at the scale of a transistor. In addition, the nanoscale modulators potentially have the advantages of small insertion loss, low power consumption, ultrahigh-speed, and easy fabrication. The successful development of this technique may lead to a significant breakthrough in on-chip optical interconnects.

**Acknowledgements:** This material is based upon work supported in part by the U.S. Army under Award No. W911NF-10-1-0153 and the National Science Foundation under Award No. ECCS-1057381.

**Figure captions**

**Figure 1**. (a) Real part and imaginary part of the dielectric constant of ITO as a function of wavelength at two different carrier concentration based on the Drude model. (b) The illustration of ENZ-slot waveguides.

**Figure 2**. (a) The plots of the transverse electric field magnitude across the ENZ-slot MIM plasmonic waveguide at $N=N_1$ and $N=N_2$, respectively. (b) The plots of the transverse electric field magnitude across the ENZ-slot dielectric waveguide at $N=N_1$ and $N=N_2$, respectively.

**Figure 3**. The electric field profiles, effective indices, and propagation loss for different ITO-slot waveguides at $N=N_1$ and $N=N_2$, respectively: (a) in a plasmonic waveguide; (b) in dielectric rib waveguide. The refractive indices of the semiconductor and $SiO_2$ are assumed to be 3.45 and 1.45, respectively.

**Figure 4**. (a) The illustration of an EO modulator embedded in a plasmonic rib waveguide. (b,c) The 3D simulation of light propagation between a plasmonic rib waveguide and the EO modulator at $N=N_1$ and $N=N_2$, respectively. (d) The illustration of an EO modulator embedded in a dielectric rib waveguide. (e,f) The 3D simulation of light propagation between a dielectric rib waveguide and the EO modulator at $N=N_1$ and $N=N_2$, respectively.

**Figures**

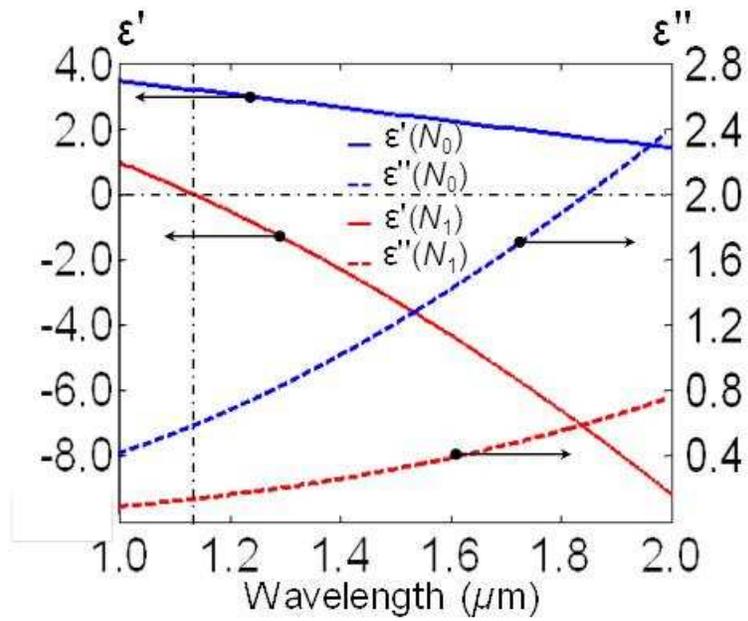

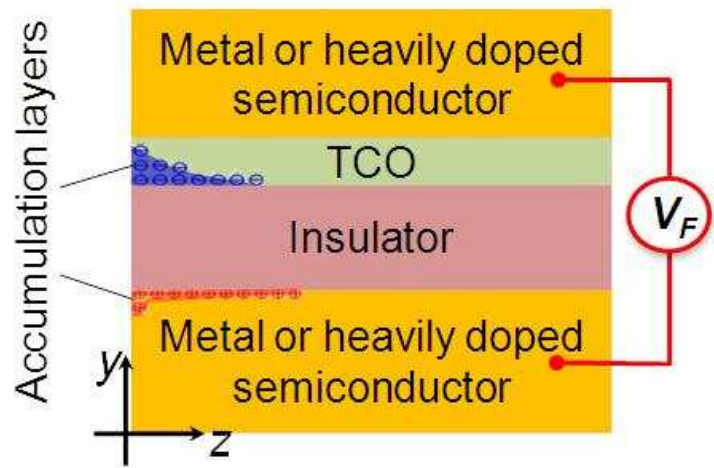

**Figure 1.**

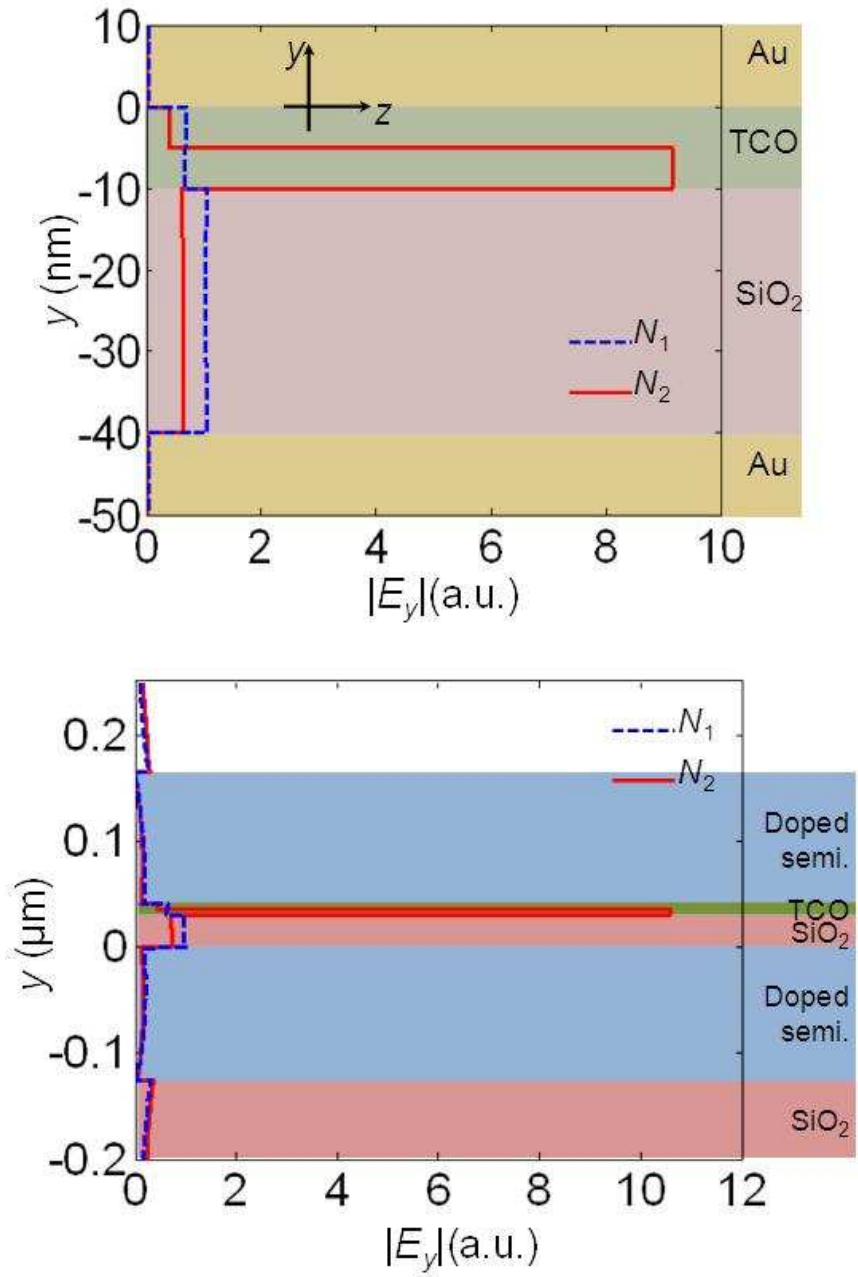

**Figure 2.**

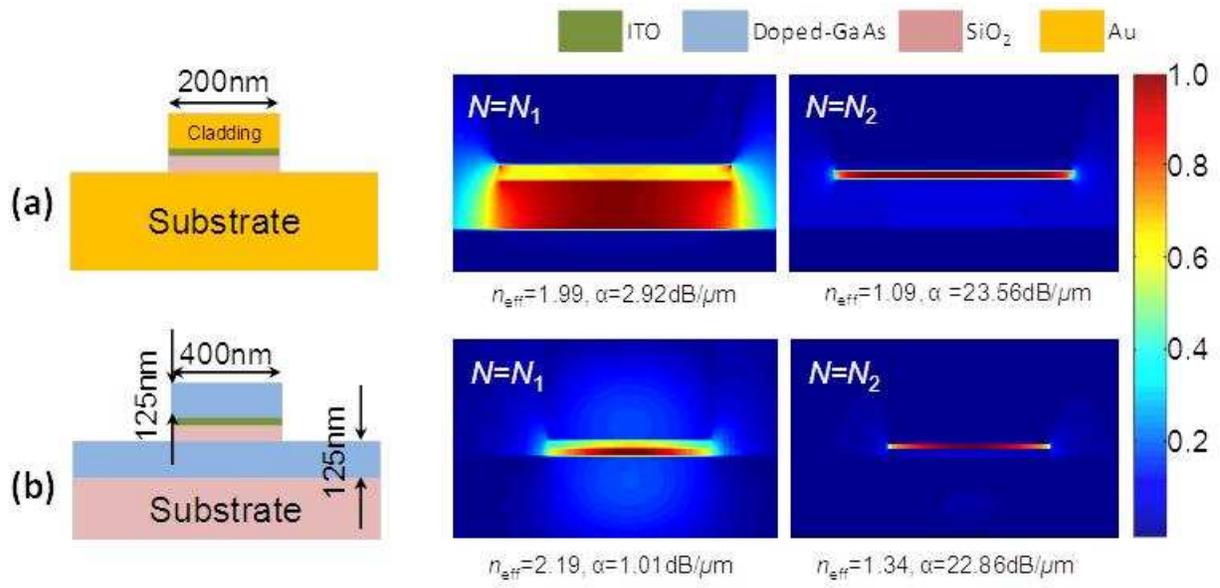

**Figure 3.**

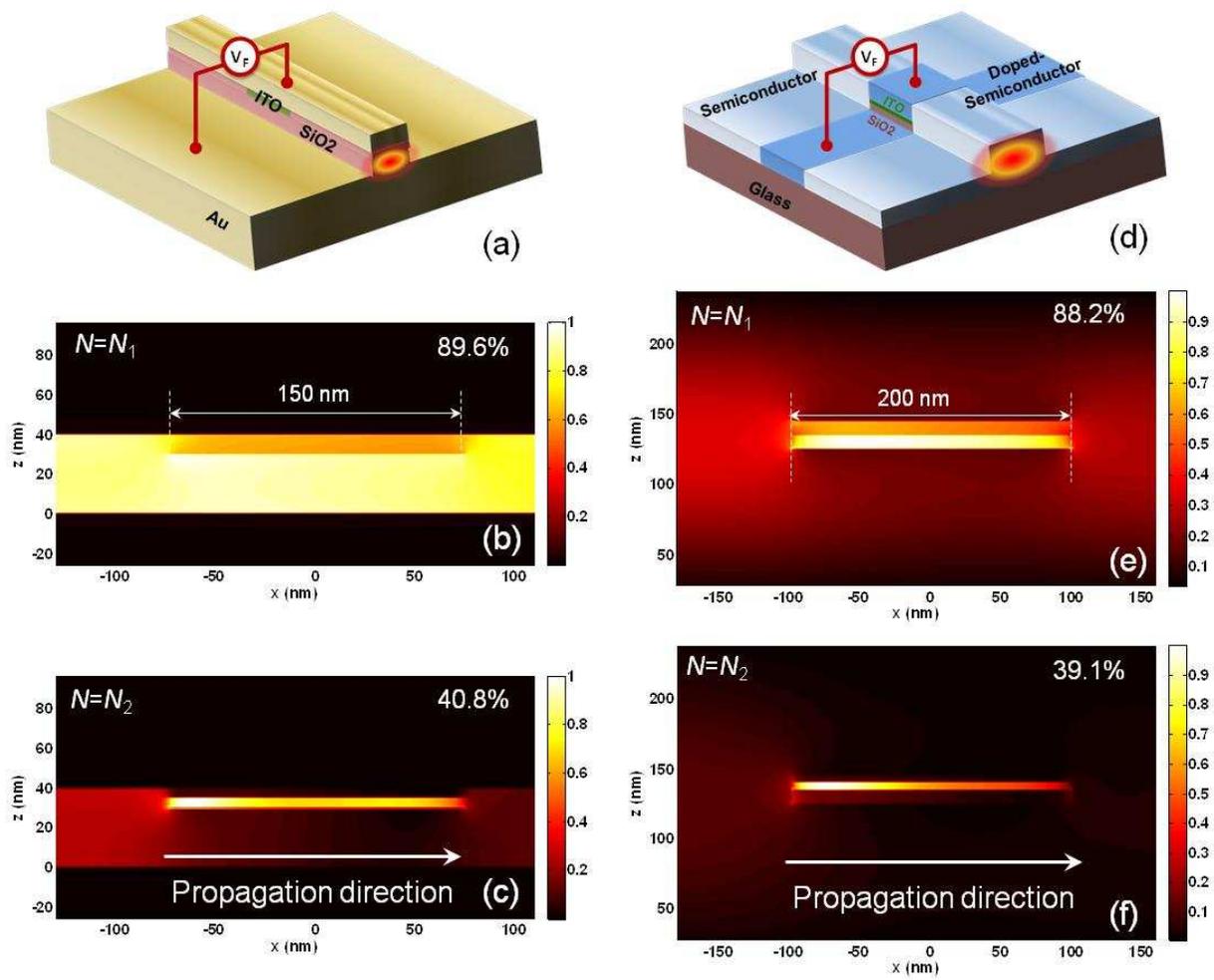

**Figure 4.**